\documentclass[twocolumn]{autart}    

\usepackage[english]{babel}                            
\usepackage{ amsmath, amssymb, amsfonts}         
\usepackage{graphicx}                                 
\usepackage{color}    
\usepackage{mathrsfs}     
\usepackage{cancel}                            
\usepackage{enumerate}
\usepackage[normalem]{ulem}
\usepackage{subcaption}
\usepackage{caption}
\usepackage{thmtools}
\usepackage{thm-restate}
\usepackage{cite}

\def\Real{\mathbb{R}}

\def\Complex{\mathbb{C}}

  \def\EE{\mathcal{E}} 
\def\GG{\mathcal{G}}

\def\RR{\boldsymbol{\rm R}} 
  \def\VV{\mathcal{V}}

\let\Pi\varPi

\newcommand{\STwoOne}[2]
{\left[\begin{smallmatrix}
{#1} \\
{#2}
\end{smallmatrix}\right]
}
\newcommand{\STwoTwo}[4]
{\left[\begin{smallmatrix}
{#1} & {#2} \\
{#3} & {#4}
\end{smallmatrix}\right]
}
\newcommand{\SOneTwo}[2]
{\left[\begin{smallmatrix}
{#1} & {#2}
\end{smallmatrix}\right]
}

\newcommand{\diag}{\mathrm{diag}}

  \newcommand{\Hinfty}{\boldsymbol{\rm H}_{\infty}}

\begin{document}

\begin{frontmatter}

\title{Steady-state analysis of networked epidemic models } % Title, preferably not more 
   \vspace{-6pt}                                               % than 10 words.

% \thanks[footnoteinfo]{This paper was not presented at any IFAC 
% meeting. Corresponding author  Lanlan Su.}

\author[1]{Sei Zhen Khong}\ead{szkhongwork@gmail.com},    
\author[2]{Lanlan Su}\ead{lanlan.su@sheffield.ac.uk}  % (ead) as shown
\address[1]{Independent researcher}  % Please supply                                              

\address[2]{Department of Automatic Control and Systems Engineering, University of Sheffield.}        % here.


\begin{abstract}
  Compartmental epidemic models with dynamics that evolve over a graph network have gained considerable importance in recent years but analysis of
  these models is in general difficult due to their complexity. In this paper, we develop two positive feedback frameworks that are applicable to the
  study of steady-state values in a wide range of compartmental epidemic models, including both group and networked
  processes. 
  In the case of a group (resp. networked) model, we show that the convergence limit of the susceptible proportion of the population (resp. the
  susceptible proportion in at least one of the subgroups) is upper bounded by the reciprocal of the basic reproduction number (BRN) of the model. The
  BRN, when it is greater than unity, thus demonstrates the level of penetration into a subpopulation by the disease. Both non-strict and strict
  bounds on the convergence limits are derived and shown to correspond to substantially distinct scenarios in the epidemic processes, one in the
  presence of the endemic state and another without. Formulae for calculating the limits are provided in the latter case. We apply the developed
  framework to examining various group and networked epidemic models commonly seen in the literature to verify the validity of our conclusions.   \vspace{-3pt}
\end{abstract}

\end{frontmatter}

\section{Introduction}   \vspace{-3pt}
Compartmental models are often applied to the study of infectious diseases in epidemiology and have enjoyed various successes~\cite{SCMV15, NPP16,
  PBB20, ZinCao21}. These models may be used to predict and analyse the spread of the diseases and potentially form the foundation on which public
health interventional control strategies are based. Epidemic models are inherently nonlinear systems, and can be significantly complex to analyse
especially if they are intended to capture more than a few compartmental features over directed networks of interacting groups or agents. Detailed
stability analysis of compartmental models in epidemiology is often restricted to the study of two or three compartments using mathematical tools such
as fixed-point theorems, Lyapunov methods, and differential geometry~\cite{LajYor76, Het00, MMZB17, YLAC21}.   \vspace{-3pt}

Numerous uses of the theory of positive systems in the study of epidemic models have been reported in the literature. Some of them are targeted at a
particular type of model, such as the networked susceptible-infected-susceptible (SIS) models in~\cite{FIST07, KBG16}, the group SIDARTHE models
in~\cite{GBBC20}, and a networked SAIR model in~\cite{SMBC22}. Others, including~\cite{DriWat02}, are applicable to general epidemic models. In
particular, a precise definition of the basic reproduction number (BRN) is presented in~\cite{DriWat02} for a general compartmental disease
transmission model, and its graphical interpretation and computation are provided in~\cite{BLD09, SisFef22}.   \vspace{-3pt}

This paper develops two positive feedback system frameworks for the steady-state analysis of a broad range of group (i.e. homogeneous mixing) and
networked (i.e. heterogeneous mixing) epidemic models. 
Importantly, we show that the BRN quantifies the `level' of penetration of the disease into at least one subgroup of the population. To be specific,
we consider two considerably distinct scenarios in epidemiology. The first predicates on the convergence of the susceptible population to the same
limit for (almost) all initial conditions, and involves marginally stable closed-loop dynamics that approximate the steady-state behaviour in the
epidemic models, whereby the existence of the endemic state is covered. This is applicable, for instance, to susceptible-infected-recovered (SIR)
models with vital birth and death dynamics. The main result is a non-strict bound on the steady-state value of the susceptible proportion of the
population in a subgroup in the network in terms of the reciprocal of the BRN. We note that in this paper we do not establish convergence in
complicated epidemic models with unique endemic equilibria --- this is an ongoing investigation in the literature. Instead, for these models we assume
convergence, and provide bounds on the steady-state values of certain subpopulations in terms of the BRNs.   \vspace{-3pt}

The second positive feedback system framework we develop allows for convergence to a limit that varies with the initial conditions, and involves
exponentially stable closed-loop dynamics for which there is no endemic state, as in the case of SIR processes without vital dynamics. The main
results are a strict bound on the steady-state value of the susceptible proportion of the population in a subgroup and formulae for computing the
steady-state values of certain compartments in the epidemic models. \vspace{-3pt} % They may be seen as generalisations of some of the results in~\cite{GBBC20}
% developed for a specific group epidemic model (SIDARTHE) to general networked epidemic models.   

The results in this paper are derived based on positive systems theory~\cite{BerPle94, TLU2013}. The recent decade has seen many developments of
positive systems theory. They include robust and scalable control of positive systems~\cite{Bri13, Ran15, ColSmi15, KBR15, KhoRan16, KaoKho18}, the
Kalman-Yakubovich-Popov lemma~\cite{TanLan11, Ran15KYP}, as well as optimal control of positive systems~\cite{CMCCB14, CMB16, DCJ18}. The rich theory
on positive systems has made compartmental models in epidemiology, which are intrinsically positive systems in that all variables stay positively
invariant over time, amenable to analysis and control via positive systems methods. \vspace{-3pt} % Interestingly, the expression of the BRN of the infection in the
% present paper is given by the spectral radius of a nonnegative matrix, which is the positively weighted DC gain of a transfer function of a stable,
% internally positive LTI system.

The paper has the following structure. First, the notation used throughout the paper is defined in the next section, alongside with the provision of
important preliminary results. The problem to be investigated is formulated in Section~\ref{sec: problem} and the main results on the convergence
limit bounds in positive feedback systems are derived in Section~\ref{sec: pos_feedback}. The latter are then applied to analysing group epidemic
models in Section~\ref{sec: group_models} and networked models in Section~\ref{sec: networked_models}. These sections are furnished with several
numerical examples that serve to affirm the validity of our main results. We note here that the developed frameworks are applicable to the study of
other, possibly more complicated, epidemic models, but we have only included a few of the commonly encountered ones in these sections for illustration
purposes. Finally, concluding remarks are provided in Section~\ref{sec: con}. \vspace{-3pt}

\section{Notation and preliminaries}   \vspace{-3pt}

\subsection{Matrix theory}\label{subsec:notation}   \vspace{-3pt}

Denote by $\Real, \Real_+, j\Real, \Complex$, $\Complex_+$, $\bar{\Complex}_-$, and $\bar{\Complex}_+$ the reals, the nonnegative reals, the imaginary
axis, the complex plane, the open right-half complex plane, the closed left-half complex plane, and the closed right-half complex plane,
respectively. Let $|\cdot|$ denote the Euclidean norm. The real part and imaginary part of $s \in \Complex$ are denoted by $\mathrm{Re}(s)$ and
$\mathrm{Im}(s)$, respectively. The $(i, j)^{\mathrm{th}}$ element of a matrix $M \in \Complex^{m \times n}$ is denoted by $m_{ij}$, and we write
$M = [m_{ij}]$. Given an $M \in \Complex^{m \times n}$ (resp.  $\Real^{m \times n}$), $M^* \in \Complex^{n \times m}$ (resp.
$M^T \in \Real^{n \times m}$) denotes its complex conjugate transpose (resp. transpose). When $m = n$, denote by $\lambda(M)$ and $\rho(M)$ the
spectrum and spectral radius of $M$, respectively. Denote by $\lambda_i(M)$, $i = 1, \ldots, n$ the eigenvalues of $M$. Given a vector
$v \in \Complex^n$, $\diag(v) \in \Complex^{n \times n}$ denotes the diagonal matrix whose diagonal entries are $v_1, \dots, v_n$.  $I_n$ denotes the
identity matrix of dimensions $n \times n$, and $1_n \in \Real^n$ the column vector of all ones. \vspace{-3pt}

Given matrices $M, N \in \Real^{m \times n}$, we write $M \geq N$ if $m_{ij} \geq n_{ij}$ for all $i$ and $j$, $M > N$ if $M \geq N$ and $M \neq N$,
and $M \gg N$ if $m_{ij} > n_{ij}$ for all $i$ and $j$. $M$ is called a nonnegative matrix if $M \geq 0$, and positive if $M \gg 0$. Given
$v, w \in \Real^n$ such that $v \gg 0$ and $w \gg 0$, let $\frac{v}{w} \in \Real^n$ denote $(\frac{v}{w})_i = \frac{v_i}{w_i}$ and
$\log (v) \in \Real^n$ be such that $\log(v)_i = \log v_i$. A square $M \in \Real^{n \times n}$ is said to be Metzler if $m_{ij} \geq 0$ for all $i \neq j$,
i.e. all its off-diagonal elements are nonnegative. $M$ is said to be Hurwitz if every eigenvalue of $M$ has strictly negative real part,
i.e. $\mathrm{Re}(\lambda_i(M)) < 0$ for every $\lambda_i(M) \in \lambda(M)$. \vspace{-3pt}

An $M \in \Real^{n \times n}$ is said to be \emph{irreducible} if there exists no permutation matrix $P$ such that
$PMP^{-1} = \STwoTwo{E}{F}{0}{G},$
where $E$ and $G$ are nontrivial square matrices, i.e. they are of dimensions greater than $0$. The following result from~\cite[Corollary
2.1.5]{BerPle94} is important for subsequent developments.   \vspace{-3pt}

\begin{lem} \label{lem: spec}
  \begin{enumerate} 
  \renewcommand{\theenumi}{\textup{(\roman{enumi})}}\renewcommand{\labelenumi}{\theenumi}
  \item If $0 \leq M \leq N$, then $\rho(M) \leq \rho(N)$; \label{item: rho_nonstrict}
 \item If $0 \leq M < N$ and $M + N$ is irreducible, then $\rho(M) < \rho(N)$. \label{item: rho_strict}
  \end{enumerate}
\end{lem}
  \vspace{-3pt}

\subsection{Graph theory}   \vspace{-3pt}

A directed graph, or digraph, is a pair $\GG = (\VV, \EE)$, where $\VV = \{1, \ldots, J\}$ is the set of nodes and $\EE \subset \VV \times \VV$,
$\EE = \{e_1, \ldots, e_m\}$ is the set of edges such that $e_k = (i, j) \in \EE$ if node $i$ is connected to node $j$, i.e. node $i$ is a neighbour
of node $j$. A graph is undirected if $(i, j) \in \EE$ then $(j, i) \in \EE$. A (directed) path on $\GG$ is an ordered set of distinct vertices
$\{n_0, n_1, \ldots, n_N\}$ such that $(n_i, n_{i + 1}) \in \EE$ for all $i \in \{0, 1, \ldots, N - 1\}$. A digraph is said to be \emph{strongly
  connected} if there is a path in each direction between each pair of nodes of the graph. Given a matrix $M \in \Real^{n \times n}$, one may associate
with it a digraph $\GG_M$ --- the graph has n nodes labeled $1, \ldots, n$ and there is an edge connecting node $i$ to node $j$ if and only if
$m_{ji} \neq 0$. Then $M$ is irreducible if and only if its associated graph $\GG_M$ is strongly connected~\cite[Theorem 2.2.7]{BerPle94}.

\subsection{Systems theory}   \vspace{-3pt}

Let $\RR$ denote the set of proper real-rational transfer functions and $\RR\Hinfty$ its subset of elements having no poles in the closed right-half
complex plane $\bar{\Complex}_+$.  For a linear time-invariant (LTI) system $G$, we denote its transfer function representation
by $\hat{G}$. For $\hat{G}\in \RR\Hinfty$, let $\|\hat{G}\|_\infty$ denote its $\Hinfty$ norm, i.e., $ \|\hat{G}\|_\infty = \sup_{\mathrm{Re}(s) > 0} \bar{\sigma}(\hat{G}(s)) < \infty$,
where $\bar{\sigma}$ denotes the largest singular value.  \vspace{-3pt}

A nonlinear system described by \vspace{-3pt}
\begin{align*} 
  \dot{x}(t) & = f(x(t), u(t)), \quad x(0) = x_0 \\
  y(t) & = h(x(t), u(t)),  
\end{align*} 
where $f(x, u)$ and $h(x, u)$ are locally Lipschitz in $(x, u)$, is said to be internally positive if $x(0) \geq 0$ and $u(t) \geq 0$ for all
$t \geq 0$, then $x(t) \geq 0$ and $y(t) \geq 0$ for all $t \geq 0$. % It is said to be externally positive if $x(0) = 0$ and $u(t) \geq 0$ for all $t
% \geq 0$, then $y(t) \geq 0$ for all $t \geq 0$. Obviously, an internally positive system is also externally positive.
An example of an internally positive system is an LTI system $G$ with
state-space realisation   \vspace{-3pt}
\begin{align} \label{eq: LTI_pos}
  \begin{split}
  \dot{x}(t) & = Ax(t) + Bu(t), \quad x(0) = x_0 \\
  y(t) & = Cx(t) + Du(t),
  \end{split}
\end{align}
where $A \in \Real^{n \times n}$ is Metzler, $B \in \Real_+^{n \times m}$, $C \in \Real_+^{p \times n}$, and $D \in \Real_+^{p \times m}$ are
nonnegative matrices; see~\cite{FarRin00}. The pair $(A, B)$ is said to be stabilisable if there exists $F$ such that $A + BF$ is Hurwitz. On the
other hand, the pair $(C, A)$ is said to be detectable if there exists $L$ such that $A + LC$ is Hurwitz; see~\cite[Chapter 3]{ZDG96}. Obviously, when
$A$ is Hurwitz, $(A, B, C)$ is stabilisable and detectable. In general, when $(A, B, C)$ is stabilisable and detectable,
$\hat{G}(s) := C(sI - A)^{-1}B + D \in \RR\Hinfty$ if and only if $A$ is Hurwitz. Likewise, the poles of $\hat{G}$ lie in $\bar{\Complex}_-$ if and
only if $\lambda(A) \subset \bar{\Complex}_-$. \vspace{-3pt}

The following important result will be used repeatedly in subsequent developments. \vspace{-3pt}

\begin{lem} \label{lem: tanaka} Consider an internally positive LTI system described by \eqref{eq: LTI_pos} with Hurwitz $A$ and $D = 0$. Given
  $K \geq 0$, it holds that $(I - K\hat{G})^{-1} \in \RR\Hinfty$ if and only if $\rho(K\hat{G}(0)) < 1$.
\end{lem}   \vspace{-3pt}
\pf
  Sufficiency follows from~\cite[Theorem 3(i)]{TLU2013}. For necessity, note that $(I - KG)^{-1} = w \mapsto u$ may be described by the LTI
  state-space model
  \begin{align*}
  \dot{x}(t) & = (A + BKC) x(t) + Bw(t) \\
  u(t) & = KC x(t) + w(t),
  \end{align*}
  which is internally positive. Suppose to the contrapositive that $\rho(K\hat{G}(0)) \geq 1$. By~\cite[Theorem 3(ii) and (iii)]{TLU2013}, this then
  implies that $(I - K\hat{G})^{-1} \notin \RR\Hinfty$.
\endpf \vspace{-3pt}

\section{Problem formulation} \label{sec: problem} 
\begin{figure}
    \centering
    \includegraphics[scale=0.5]{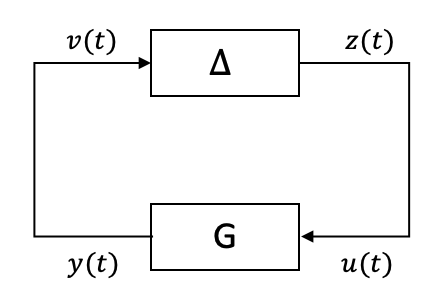}
    \centering \caption{The feedback system $[G,\Delta]$}
    \label{fig:GDelta}
  \end{figure}

\vspace{-3pt}
Let the nonlinear system $\Delta$ be an internally positive system described by 
\begin{align} \label{eq: nonlinear}
  \begin{split}
  \dot{s}(t) & = f(s(t), v(t)), \quad s(0) = s_0 \\
  z(t) & = M_1 \diag(s(t)) M_2 v(t),
  \end{split}
\end{align}
where $f(s, v)$ is locally Lipschitz in $(s, v)$, $s(t) \in \Real^{n_s}$, $M_1 > 0$, and $M_2 > 0$. Next, denote by $G$ an internally positive
LTI system   \vspace{-3pt}
\begin{align} \label{eq: LTI}
  \begin{split}
  \dot{x}(t) & = Ax(t) + Bu(t), \quad x(0) = x_0 \\
   y(t) & = Cx(t),
  \end{split}
\end{align}
where $A \in \Real^{n \times n}$ is Metzler and Hurwitz, $B > 0$ and $C > 0$ are nonzero nonnegative matrices. Consider the positive feedback
interconnection of $G$ and the internally positive system $\Delta$ in \eqref{eq: nonlinear} in which   \vspace{-3pt}
\begin{align} \label{eq: CL}
  v = y, \quad u = z.
\end{align}
We write the resultant feedback system modeled by \eqref{eq: nonlinear}, \eqref{eq: LTI}, and
\eqref{eq: CL} as $[G, \Delta]$, which is depicted in Figure \ref{fig:GDelta}.
In real applications, one may not initialise the system at arbitrary initial conditions but only those that are of
significance. Let $\mathscr{I}$ denote the set of initial conditions $(s(0), x(0))$ of interest. \vspace{-3pt}

The objective of this paper is as follows. Suppose for all $(s(0), x(0)) \in \mathscr{I}$, it holds that $\lim_{t \to \infty} s(t)$ exists. Find the
limit or an upper bound on the limit in the case where $s(t)$ is a scalar, and an upper bound on an entry of $s(t)$ otherwise. We achieve the
objective above in the next section using mathematical tools from positive systems theory. \vspace{-3pt}

When applied to epidemic models, the $s(t)$ above is taken to denote the proportions of the populations in different groups that are susceptible to a
contagious disease. An upper bound on an element in $\lim_{t \to \infty} s(t)$ then indicates the level of penetration of the disease into the
population in the corresponding group. \vspace{-3pt}

\section{Positive feedback systems} \label{sec: pos_feedback}   \vspace{-3pt}

\subsection{Non-strict bound on equilibrium}   \vspace{-3pt}

First, a couple of assumptions are stated.

\begin{assum} \label{as: equi} If $f(\bar{x}, \bar{u}) = 0$, then for sufficiently small $\epsilon > 0$, it holds that $u \neq \bar{u}$ and
  $|u - \bar{u}| < \epsilon$ imply $f(\bar{x}, u) \neq 0$.
\end{assum}   \vspace{-3pt}

\begin{assum} \label{as: initial_cond} If $(s, x) \in \mathscr{I}$, then for all $z > 0$, there exists $\tau > 0$ such that $(s, \tau z) \in \mathscr{I}$.
\end{assum} \vspace{-3pt}

\begin{thm} \label{thm: bound} Consider the feedback system $[G, \Delta]$ described by \eqref{eq: nonlinear}, \eqref{eq: LTI}, and \eqref{eq: CL}. Suppose Assumptions~\ref{as: equi} and~\ref{as: initial_cond} hold, $-M_2CA^{-1}BM_1$ is
  irreducible, and for all initial conditions $(s(0), x(0)) \in \mathscr{I}$, $\lim_{t \to \infty} s(t) = \bar{s}$. Then there exists $i \in \{1, \ldots, n_s\}$ such that   \vspace{-3pt}
\[
  \bar{s}_i \leq \frac{1}{\rho(M_2\hat{G}(0)M_1)} = \frac{1}{\rho(-M_2CA^{-1}BM_1)}.
\]
Moreover, if $\lim_{t \to \infty} x(t) = \bar{x} > 0$,
then there exists $i$ such that $\bar{s}_i < \frac{1}{\rho(M_2\hat{G}(0)M_1)}$ 
if and only if there exists $j \neq i$ such that
$\bar{s}_j > \frac{1}{\rho(M_2\hat{G}(0)M_1)}$. In other words, if $n_s = 1$, then $\bar{s} = \frac{1}{\rho(M_2\hat{G}(0)M_1)}$.
\end{thm}
 \vspace{-3pt}
\pf
  As $\lim_{t \to \infty} s(t) = \bar{s}$, it must hold that
$f(\bar{s}, v(t)) = f(\bar{s}, y(t)) \to 0$
as $t \to \infty$, which implies by Assumption~\ref{as: equi} that $y(t) \to \bar{y}$ for some $\bar{y}$, whereby $x(t) \to \bar{x}$ for some
$\bar{x}$ because $(C, A)$ is detectable.

Since $s(t) \to \bar{s}$, it follows that for sufficiently large $T > 0$, by approximating $\Delta$ by the constant gain   \vspace{-3pt}
\[
  \tilde{\Delta} = v \mapsto z : z(t) = M_1 \diag(\bar{s}) M_2 v(t) =: Kv(t)
\]
for $t \geq T$, the closed-loop system $[G, \tilde{\Delta}]$ described by   \vspace{-3pt}
\begin{align} \label{eq: autonomous_CL_LTI}
\dot{x}(t) = (A + BKC) x(t) 
\end{align}
is a close approximation of the dynamics in $[G, \Delta]$ for $t \geq T$. The fact that $x(t) \to \bar{x}$ then implies that
$\lambda(A + BKC) \subset \bar{\Complex}_-$. \vspace{-3pt}

To see this, suppose $\lambda(A + BKC) \cap \Complex_+ \neq \emptyset$. Since $A + BKC$ is Metzler, we can write it as $A + BKC = M - b I$ for some
$b > 0$ and $M \geq 0$. By the Krein-Rutman theorem for nonnegative matrices~\cite[Theorem 2.1.1]{BerPle94}, there then exists $z > 0$ such that
$(A + BKC)z = (M - b I)z = \kappa z$ for some $\kappa > 0$. Since $(\bar{s}, \tau z) \in \mathscr{I}$ for some $\tau > 0$ by Assumption~\ref{as:
  initial_cond}, setting $(s(0), x(0)) = (\bar{s}, \tau z) \in \mathscr{I}$ in \eqref{eq: nonlinear}, \eqref{eq: LTI}, and \eqref{eq: CL} then yields
that $|x(t)| \to \infty$ in \eqref{eq: autonomous_CL_LTI}, leading to a contradiction to $x(t) \to \bar{x}$. Therefore, it must hold that   \vspace{-3pt}
\begin{align} \label{eq: CLHP}
  \lambda(A + BKC) \subset \bar{\Complex}_-.
\end{align}

Notice that \eqref{eq: CLHP} implies $\lambda(A - \alpha I + BKC) \subset \Complex_-$
for all $\alpha > 0$. Now consider   \vspace{-3pt}
\begin{align*} 
  \begin{split}
  \dot{x}(t) & = (A -\alpha I + BKC) x(t) + Bw(t) \\
  u(t) & = KC x(t) + w(t),
  \end{split}
\end{align*}
which describes the closed-loop system
$(I - \tilde{\Delta}G_\alpha)^{-1} = w \mapsto u$,
where $\hat{G}_\alpha(s): = C(sI - (A - \alpha I))^{-1}B$. Evidently, the LTI system above is internally positive. Since $A - \alpha I + BKC$ is
Hurwitz, it follows that $(I - \tilde{\Delta}\hat{G}_\alpha(s))^{-1} \in \RR\Hinfty$.
From~Lemma~\ref{lem: tanaka}, $(I - \tilde{\Delta}\hat{G}_\alpha(s))^{-1} \in \RR\Hinfty$ if
and only if
\begin{equation}\label{eq:DCgainless1}
  \rho(\tilde{\Delta}\hat{G}_\alpha(0)) < 1.
\end{equation}
By continuity, as $\alpha \to 0$, we have $\rho(\tilde{\Delta}\hat{G}(0)) \leq 1$, where $\hat{G}(s) = C(sI - A)^{-1}B$. \vspace{-3pt}

Recall from~\cite[Theorem 6.2.3]{BerPle94}\cite[Proposition 1]{Ran15} that the Metzler matrix $A$ is Hurwitz if and only if $-A^{-1} \geq 0$. Thus,
$-M_2 CA^{-1}B M_1 \geq 0$. Observe that   \vspace{-3pt}
\begin{align} \label{eq: rho_M}
  \begin{split}
  \rho(\tilde{\Delta}\hat{G}(0)) & = \rho(- M_1 \diag(\bar{s}) M_2  CA^{-1}B) \\
  & = \rho(- \diag(\bar{s}) M_2  CA^{-1}B M_1).
  \end{split}
\end{align}
If $\bar{s} = \frac{1}{\rho(-M_2CA^{-1}BM_1)} 1_{n_s}$, then clearly $\rho(\tilde{\Delta}\hat{G}(0)) = 1$. By hypothesis, $-M_2CA^{-1}BM_1 \geq 0$ is irreducible. Since $\diag(\bar{s}) \geq 0$, it follows that
$\diag(\bar{s}) - M_2CA^{-1}BM_1$
is irreducible~\cite[Corollary 2.1.10(a)]{BerPle94}. Therefore, by Lemma~\ref{lem: spec}\ref{item: rho_strict}, if
$\bar{s} > \frac{1}{\rho(-M_2CA^{-1}BM_1)} 1_{n_s}$, then $\rho(\tilde{\Delta}\hat{G}(0)) > 1$. In other words, $\rho(\tilde{\Delta}\hat{G}(0)) \leq 1$
implies that there exists $i$ such that   \vspace{-3pt}
\begin{align} \label{eq: bound_s}
  \bar{s}_i \leq \frac{1}{\rho(-M_2CA^{-1}BM_1)}.
\end{align}
This completes the proof for the first claim.
For the second claim, note that $x(t) \to \bar{x} > 0$ in \eqref{eq: autonomous_CL_LTI} implies that $0 \in \lambda(A + BKC)$. Since
$\rho(\tilde{\Delta}\hat{G}(0)) \leq 1$ holds as shown above, $0 \in \lambda(A + BKC)$ only if $\rho(\tilde{\Delta}\hat{G}(0)) = 1$. To see this,
observe that $\rho(\tilde{\Delta}\hat{G}(0)) < 1$ would imply that $(I - \tilde{\Delta}\hat{G})^{-1} \in \RR\Hinfty$ by Lemma~\ref{lem: tanaka}, where
$(I - \tilde{\Delta} G)^{-1}$ is the internally positive LTI system described by   \vspace{-3pt}
\begin{align} \label{eq: CL_LTI}
  \begin{split}
  \dot{x}(t) & = (A + BKC) x(t) + Bw(t) \\
  u(t) & = KC x(t) + w(t).
  \end{split}
\end{align}
This would in turn imply that $A + BKC$ is Hurwitz because $(A + BKC, B, KC)$ is stabilisable and detectable, which follows from the fact that $A$ is
Hurwitz. This leads to a contradiction. \vspace{-3pt}

By the same reasoning leading to $\bar{s} > \frac{1}{\rho(-M_2CA^{-1}BM_1)} 1_{n_s}$ implies $\rho(\tilde{\Delta}\hat{G}(0)) > 1$ above, it may be
shown similarly that $\bar{s} < \frac{1}{\rho(-M_2CA^{-1}BM_1)} 1_{n_s}$ implies $\rho(\tilde{\Delta}\hat{G}(0)) < 1$. Therefore, since
$\rho(\tilde{\Delta}\hat{G}(0)) = 1$, it holds that there exists $i$ such that
$\bar{s}_i < \frac{1}{\rho(-M_2CA^{-1}BM_1)} $
if and only if there exists $j \neq i$ such that
$\bar{s}_j > \frac{1}{\rho(-M_2CA^{-1}BM_1)}$,
as required.
\endpf
\vspace{-3pt}

The following result is of independent interest and significance. It shows that $\rho(\diag(\bar{s})M_2\hat{G}(0)M_1) \leq 1$ in the proof of
Theorem~\ref{thm: bound} is necessary and sufficient for the eigenvalues of the state matrix $A + BKC$ of the approximating LTI closed-loop system
$(I - \tilde{\Delta} G)^{-1} = w \mapsto u$ described by \eqref{eq: CL_LTI} to lie in $\bar{\Complex}_-$. \vspace{-3pt}

\begin{thm} \label{thm: bound_iff} Suppose $-M_2CA^{-1}BM_1$ is irreducible.  Then   \vspace{-3pt}
  \begin{align} \label{eq: bound}
    \begin{split}
    & \rho(\diag(\bar{s})M_2\hat{G}(0)M_1) \\
    = \; & \rho(-\diag(\bar{s})M_2CA^{-1}BM_1) \leq 1
    \end{split}
\end{align}
if and only if $\lambda(A + BKC) \subset \bar{\Complex}_-$, where
$K := M_1 \diag(\bar{s}) M_2$. Furthermore, \eqref{eq: bound} holds with equality if and only if $\lambda(A + BKC) \subset \bar{\Complex}_-$ and $0 \in \lambda(A + BKC)$.
\end{thm}
\vspace{-3pt}

\pf
  We only show the first part of the theorem since the second part may be proven similarly. To be specific, sufficiency may be established as in the
  proof of Theorem~\ref{thm: bound}, starting from \eqref{eq: CLHP} and leading to the conclusion in \eqref{eq:DCgainless1}. To show necessity, note that
  by~Lemma~\ref{lem: tanaka}, $\rho(K\hat{G}(0)) \leq 1$
implies that $(I - \alpha K\hat{G})^{-1} \in \RR\Hinfty$ for all $\alpha \in [0, 1)$. It thus follows from continuity that the poles of
  $(I - K\hat{G})^{-1}$ lie in $\bar{\Complex}_-$. Recall that a state-space realisation of $(I - K G)^{-1}$ is given by
  \eqref{eq: CL_LTI}. Because $A$ is Hurwitz, $(A + BKC, B)$ is stabilisable and $(KC, A + BKC)$ is detectable. Altogether, this means
  $\lambda(A + BKC) \subset \bar{\Complex}_-$. Noting that \eqref{eq: bound} is equivalent to $\rho(K\hat{G}(0)) \leq 1$ then completes
  the proof.
  \endpf
  \vspace{-3pt}

\vspace{-3pt}
\subsection{Strict bound on equilibrium} \vspace{-3pt}

The next result, Theorem~\ref{thm: bound_strict}, shows that if Assumption~\ref{as: equi0} is used in lieu of Assumption~\ref{as: equi}, and
Assumption~\ref{as: initial_cond} is strengthened to Assumption~\ref{as: initial_cond_strong} below, then the bound in Theorem~\ref{thm: bound} holds
with $\leq$ replaced by $<$ even when the irreducibility assumption is dropped and the steady-state value varies with initial conditions. It is
applicable to general multi-input-multi-output (MIMO) systems and a generalisation of \cite[Proposition 2]{GBBC20}, which was developed for a specific
single-input-single-output (SISO) system called the SIDARTHE model.  \vspace{-3pt}

\begin{assum} \label{as: equi0} If $\bar{x} \neq 0$ and $f(\bar{x}, \bar{u}) = 0$, then $\bar{u} = 0$.
\end{assum} \vspace{-3pt}

\begin{assum} \label{as: initial_cond_strong} If $(s, x) \in \mathscr{I}$, then for all $z > 0$, $(s, \tau z) \in \mathscr{I}$ for sufficiently small
  $\tau > 0$.
\end{assum} \vspace{-3pt}

\begin{thm} \label{thm: bound_strict} Consider the feedback system $[G, \Delta]$ described by \eqref{eq: nonlinear}, \eqref{eq: LTI}, and \eqref{eq: CL}. Suppose Assumptions~\ref{as: equi0} and~\ref{as: initial_cond_strong} hold, and for all initial conditions
  $(s(0), x(0)) \in \mathscr{I}$,
$\bar{s}(s(0), x(0)) := \lim_{t \to \infty} s(t)$
is well defined and continuous in $x(0)$. Then
$x(t) \to 0$
and there exists $i \in \{1, \ldots, n_s\}$ such that \vspace{-3pt}
\begin{align} \label{eq: strict_bound_on_s}
  \begin{split}
  \bar{s}_i(s(0), x(0)) & < \frac{1}{\rho(M_2\hat{G}(0)M_1)} \\
  & = \frac{1}{\rho(-M_2CA^{-1}BM_1)}.
  \end{split}
\end{align}
\end{thm}
\vspace{-3pt}

\pf
 Given $(s(0), x(0)) \in \mathscr{I}$, since $\bar{s}(s(0), x(0))$ exists, it must
  hold that \vspace{-3pt}
\[
  f(\bar{s}(s(0), x(0)), v(t)) = f(\bar{s}(s(0), x(0)), y(t)) \to 0
\]
as $t \to \infty$. Moreover, if $\bar{s}(s(0), x(0)) > 0$, then this implies by Assumption~\ref{as: equi0} that $y(t) \to 0$. Since $(C, A)$ is detectable, it
follows that $x(t) \to 0$. For sufficiently large $T > 0$, by approximating $\Delta$ by the
constant gain \vspace{-3pt}
\[
  \tilde{\Delta} = v \mapsto z : z(t) = M_1 \diag(\bar{s}(s(0), x(0))) M_2 v(t) =: Kv(t)
\]
for $t \geq T$, the closed-loop system $[G, \tilde{\Delta}]$ described by \vspace{-3pt}
\begin{align} \label{eq: autonomous_CL_LTI_strict}
\dot{x}(t) = (A + BKC) x(t)
\end{align}
is a close approximation of the dynamics in $[G, \Delta]$ for $t \geq T$. Note that if $\bar{s}(s(0), x(0)) = 0$, then trivially $K = 0$ and $A + BKC = A$ is
Hurwitz, in which case $x(t) \to 0$. \vspace{-3pt}

Thus, consider $\bar{s}(s(0), x(0)) > 0$. Since $A + BKC$ is Metzler, the fact that $x(t) \to 0$ then implies that $A + BKC$ is Hurwitz. To see this,
suppose $\lambda(A + BKC) \cap \bar{\Complex}_+ \neq \emptyset$. Write $A + BKC = M - b I$ for some $b > 0$ and $M \geq 0$. By the Krein-Rutman
theorem for nonnegative matrices~\cite[Theorem 2.1.1]{BerPle94}, there then exists $z > 0$ such that $(A + BKC)z = (M - b I)z = \kappa z$ for some
$\kappa \geq 0$. By Assumption~\ref{as: initial_cond_strong}, $(\bar{s}, \tau z) \in \mathscr{I}$ for sufficiently small $\tau > 0$. Setting
$(s(0), x(0)) = (\bar{s}, \tau z) \in \mathscr{I}$ in \eqref{eq: nonlinear}, \eqref{eq: LTI}, and \eqref{eq: CL} for a sufficiently small $\tau > 0$
and exploiting continuity of $\bar{s}$ in $x(0)$ then yields in \eqref{eq: autonomous_CL_LTI_strict} a sufficiently small perturbation on $K$ and
either $\lim_{t \to \infty} x(t) = \tau z > 0$ (if $\kappa = 0$) or $|x(t)| \to \infty$ (if $\kappa > 0$). This leads to a contradiction to
$x(t) \to 0$. As such, $A + BKC$ must be Hurwitz, whereby $x(t) \to 0$. \vspace{-3pt}
\begin{figure}
    \centering
    \includegraphics[scale=0.5]{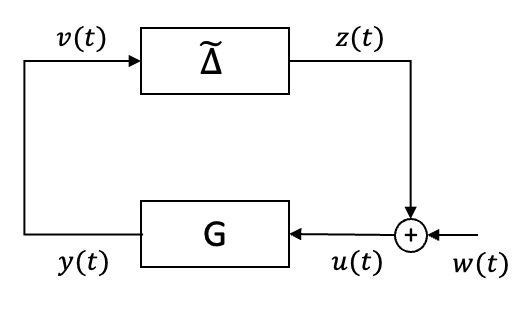}
    \centering \caption{The feedback system $[G,\tilde{\Delta}]$}
    \label{fig:GDeltaT}
  \end{figure}

  Now, define
\begin{align} \label{eq: CL_LTI_strict}
  \begin{split}
  \dot{x}(t) & = (A + BKC) x(t) + Bw(t) \\
  u(t) & = KC x(t) + w(t),
  \end{split}
\end{align}
which is internally positive and describes the closed-loop system in Figure \ref{fig:GDeltaT}, i.e., $(I - \tilde{\Delta}G)^{-1} = w \mapsto u$. \vspace{-3pt}

Hurwitzness of $A + BKC$ implies that
$(I - \tilde{\Delta}\hat{G})^{-1} \in \RR\Hinfty$,
where $\hat{G}(s) = C(sI - A)^{-1}B$. By~Lemma~\ref{lem: tanaka}, $(I - \tilde{\Delta}\hat{G})^{-1} \in \RR\Hinfty$ if and only if
$\rho(\tilde{\Delta}\hat{G}(0)) < 1$.

Recalling \eqref{eq: rho_M}, if $\bar{s} = \frac{1}{\rho(-M_2CA^{-1}BM_1)} 1_{n_s}$, then clearly
$\rho(\tilde{\Delta}\hat{G}(0)) = 1$. Furthermore, by Lemma~\ref{lem: spec}\ref{item: rho_nonstrict}, if
$\bar{s} \geq \frac{1}{\rho(-M_2CA^{-1}BM_1)} 1_{n_s}$, then $\rho(\tilde{\Delta}\hat{G}(0)) \geq 1$. Hence, $\rho(\tilde{\Delta}\hat{G}(0)) < 1$
implies that there exists $i$ such that
$\bar{s}_i < \frac{1}{\rho(-M_2CA^{-1}BM_1)}$,
as claimed.
\endpf
\vspace{-3pt}

\begin{rem}
  A crucial difference between Theorems~\ref{thm: bound} and~\ref{thm: bound_strict} is that the limit $\bar{s}$ in the former is the same for all
  initial conditions in $\mathscr{I}$, whereas in the latter, the limit $\bar{s}(s(0), x(0))$ is dependent on the initial conditions
  $(s(0), x(0)) \in \mathscr{I}$.
\end{rem} \vspace{-3pt}

The following result is a counterpart to Theorem~\ref{thm: bound_iff}. It shows that $\rho(\diag(\bar{s}) M_2\hat{G}(0) M_1) < 1$ in the proof of
Theorem~\ref{thm: bound_strict} is necessary and sufficient for the internal stability of the approximating LTI closed-loop system
$(I - \tilde{\Delta} G)^{-1}$ described by \eqref{eq: CL_LTI_strict}. The result is a MIMO generalisation of the SISO result in \cite[Proposition 1]{GBBC20}. \vspace{-3pt}

\begin{thm} \label{thm: bound_iff_strict} It holds that \vspace{-3pt}
  \begin{align} \label{eq: bound_strict}
    \begin{split}
      & \rho(\diag(\bar{s}) M_2\hat{G}(0) M_1) \\
      = \; & \rho(-\diag(\bar{s})M_2CA^{-1}BM_1) < 1
  \end{split}
\end{align}
if and only if $A + BKC$ is Hurwitz, where $K := M_1 \diag(\bar{s}) M_2$. 
\end{thm}
\vspace{-3pt}

\pf
  Note that because $A$ is Hurwitz, $(A + BKC, B)$ is stabilisable and $(KC, A + BKC)$ is detectable. As such, $(I - \tilde{\Delta}\hat{G})^{-1}$ with
  state-space realisation given in \eqref{eq: CL_LTI_strict} is an element of $\RR\Hinfty$ if and only if $A + BKC$ is Hurwitz. Since the LTI system
  in \eqref{eq: CL_LTI_strict} is internally positive, it follows by~Lemma~\ref{lem: tanaka} that $(I - K\hat{G})^{-1} \in \RR\Hinfty$ if
  and only if
$ 
    \rho(K\hat{G}(0)) < 1.
$ 
  Noting that \eqref{eq: bound_strict} is equivalent to $\rho(K\hat{G}(0)) < 1$ then completes the proof.
\endpf
\vspace{-3pt}

\subsection{Equilibria for specific models} \vspace{-3pt}

The subsequent result shows that if $f$ in \eqref{eq: nonlinear} takes a specific form, then a characterisation of the equilibrium may be obtained. It
may be applied to general MIMO systems and is a generalisation of \cite[Proposition 3]{GBBC20}, which targets a specific SISO system. \vspace{-3pt}

\begin{thm} \label{thm: s_value} Consider the feedback system $[G, \Delta]$ described by \eqref{eq: nonlinear}, \eqref{eq: LTI}, and \eqref{eq:
    CL}. Suppose Assumptions~\ref{as: equi0} and~\ref{as: initial_cond_strong} hold, $M_1 = I$, $f(s, v) = - \diag(s) M_2 v$,
and for $(s(0), x(0)) \in \mathscr{I}$, $s(0) \gg 0$, it holds $s(t)\gg 0,\forall t\ge 0$ and  $\lim_{t \to \infty} s(t) = \bar{s} \gg 0$. 
Then $\bar{s}$ satisfies \vspace{-3pt}
\[
 M_2CA^{-1} x(0) = \log\left(\diag(s(0))^{-1}\bar{s}\right) + M_2CA^{-1} B (\bar{s} - s(0)).
\]
Furthermore, if $\dot{r}(t) = Mx(t)$ for some $M > 0$ and $\displaystyle\lim_{t \to \infty} r(t) = \bar{r}$, then $\bar{r}$ satisfies \vspace{-3pt}
\[
    -MA^{-1}x(0) = \bar{r} - r(0) - MA^{-1}B(\bar{s} - s(0)).
\]
\end{thm}
\vspace{-3pt}

\pf
 First, by the same arguments as in the proof of Theorem~\ref{thm: bound_strict}, the existence of $\lim_{t \to \infty} s(t)$ implies
 that $x(t) \to 0$. By hypothesis,
$
   \dot{s}(t) = - \diag(s(t)) M_2 v(t),
$ 
 whereby
$\frac{d}{dt} \log (s(t)) = \diag(s(t))^{-1}\dot{s}(t)= - M_2 v(t)$.
Since $u(t)=z(t)=\diag(s(t))M_2v(t)$, it follows that
$u(t)=-\dot{s}(t)$.
From \eqref{eq: nonlinear}, \eqref{eq: LTI}, and \eqref{eq: CL}, we have \vspace{-3pt}
 \begin{align} \label{eq: x_dot}
   \begin{split}
   \int_0^\infty \dot{x}(t) \, dt & = x(\infty) - x(0) = -x(0) \\
                                  & = A \int_0^\infty x(t) \, dt + B \int_0^\infty u(t) \, dt \\
                                  & = A \int_0^\infty x(t) \, dt - B \int_0^\infty \dot{s}(t) \, dt.
                                \end{split}                           
 \end{align}
 Premultiplying the equation above by $-M_2CA^{-1}$ yields \vspace{-3pt}
  \begin{align*}
 & M_2CA^{-1} x(0)\\
= &-M_2 \int_0^\infty v(t) dt + M_2CA^{-1} B \int_0^\infty \dot{s}(t)  dt\\
= &\int_0^\infty \frac{d}{dt} \log (s(t))  dt + M_2CA^{-1} B \int_0^\infty \dot{s}(t)  dt\\
 = & \log\left(\diag(s(0))^{-1}\bar{s}\right) + M_2CA^{-1} B (\bar{s} - s(0)).
  \end{align*}
Consider now $\dot{r}(t) = Mx(t)$. Premultiplying \eqref{eq: x_dot} by $MA^{-1}$ yields that
$-MA^{-1}x(0)  = \int_0^\infty \dot{r}(t) \, dt - MA^{-1} B \int_0^\infty \dot{s}(t) \, dt  = \bar{r} - r(0) - MA^{-1}B(\bar{s} - s(0))$,
as required. \vspace{-3pt}
\endpf

\subsection{Application to epidemic models} \vspace{-3pt}

The feedback system modeled by \eqref{eq: nonlinear}, \eqref{eq: LTI}, and \eqref{eq: CL} can be applied to various group and epidemic models by
taking $s_i(t)$ as the proportion of a susceptible population and $x_i(t)$ the proportion of a population that belongs to disease compartments,
where there are $N$ populations in total and $i \in \{1, \ldots, N\}$. By substituting the output of \eqref{eq: nonlinear} into \eqref{eq: LTI}, one
obtains \vspace{-3pt}
\[
 \dot{x} = Ax + B M_1 \diag(s(t)) M_2 Cx(t).
\]
Linearising the model around the disease-free equilibrium $s^* = [1, \ldots, 1]^T$, $x^* = 0$ then yields $\dot{x} = Ax + B M_1M_2 Cx(t)$.
Under certain assumptions, \cite[Theorem. 2]{DriWat02} shows that with
$R_0 := \rho(BM_1M_2CA^{-1})$,
the disease-free equilibrium of the epidemic model is locally asymptotically stable if $R_0 < 1$ and unstable if $R_0 > 1$; see
also~\cite{BraCas12}. Notice that \vspace{-3pt}
\[
\frac{1}{R_0} = \frac{1}{\rho(BM_1M_2CA^{-1})} = \frac{1}{\rho(-M_2CA^{-1}BM_1)},
\]
which is the derived upper bound on the steady-state value of $s(t)$ in Theorems~\ref{thm: bound} and \ref{thm: bound_strict}. The value $R_0$ is of
significant importance in the study of convergence to equilibria in epidemic models, and is known as the \emph{basic reproduction number} (BRN)
$R_0$. It captures the average spreadability of communicable diseases and is considered a fundamental threshold in epidemiology. More specifically, it
represents the expected number of secondary infections arising from an infected individual, i.e. the average number of persons to which an infected
person can pass the disease. \vspace{-3pt}

Of particular interest is the case where $R_0 > 1$, i.e. the disease-free equilibrium is unstable. Under considerably different circumstances,
each of Theorem~\ref{thm: bound} and Theorem~\ref{thm: bound_strict} provides an upper bound on an entry in $\lim_{t \to \infty} s(t)$ in the form of
$\frac{1}{R_0}$. This indicates the level of penetration of the disease into at least one subpopulation. Furthermore, Theorem~\ref{thm: bound} allows
for $x(t)$ to converge to a nonzero value, which in epidemiology corresponds to the endemic state. On the contrary, when the suppositions of
Theorem~\ref{thm: bound_strict} are satisfied, it must hold that $x(t) \to 0$, meaning that there is no endemic state, i.e. the entire population that has
caught the disease has either recovered or succumbed to the disease. \vspace{-3pt}

\section{Group epidemic models} \label{sec: group_models}\vspace{-3pt}

In this section and the next, we apply the main results developed in the previous section to the analysis of the steady-state values of the spread
dynamics in epidemic models. The include both group and networked models of compartmental form found in the literature. The present section focuses on
group models, whereas the next section is dedicated to studying networked models. \vspace{-3pt}

It is noteworthy that we do not establish convergence in the epidemic models studied in this and the next sections. The susceptible populations in
some of the models under study are bound to converge due to the monotone convergence theorem, while for some others convergence has been established
in the literature. For the more complicated models for which convergence analysis of endemic equilibria has not been completed, we simply  assume convergence
and apply our results to obtain bounds on the susceptible subpopulation. While convergence in these models have not been formally established, it has been
observed in simulations, including those provided in this paper.  \vspace{-3pt}

\subsection{SIS models} \vspace{-3pt}

The SIS model introduced in~\cite{KerMcK32} is given by:\vspace{-3pt}
\begin{align*}
  \dot{S}(t) &= -\beta S(t) I(t) + \gamma I(t) \\
  \dot{I}(t) & = \beta S(t) I(t) - \gamma I(t),
\end{align*}
where $S(t)$ denotes the proportion of the population that is susceptible to a disease at time $t$, $I$ the proportion that is infected, $\beta > 0$ the rate of
infection, or the contact between susceptible and infected compartments of the population, and $\gamma > 0$ the rate of healing or recovery of the
infected populace. The entire population is normalised to $1$. Note that if $S(0) + I(0) = 1$, then $S(t) + I(t) = 1$ for all $t \geq 0$ because
$\dot{S}(t) + \dot{I}(t) = 0$, i.e. the total mass of the population is preserved over time. Thus, the set of initial conditions of interest is
$\mathscr{I} := \{(S, I) : S \geq 0, I > 0, S + I = 1\}$, which satisfies Assumption~\ref{as: initial_cond}. \vspace{-3pt}

It is well known \cite{KerMcK32} that the BRN for an SIS model is $R_0 = \frac{\beta}{\gamma}$, i.e. the ratio of the infection rate to recovery rate. When $R_0 \leq 1$,
the disease-free state ($S = 1$, $I = 0$) is a globally asymptotically stable equilibrium. On the other hand, when $R_0 > 1$, the endemic state
($S = \frac{\gamma}{\beta}$, $I = 1 - \frac{\gamma}{\beta}$) is almost globally asymptotically stable, with convergence guaranteed for all initial
conditions $S(0) + I(0) = 1$ except when $I(0) = 0$.
Define LTI system $G$ as in \eqref{eq: LTI} with $x(t) := I(t)$, $A := -\gamma$, $B := \beta$, $C := 1$, and nonlinear system $\Delta$ as in
\eqref{eq: nonlinear} with $s(t) := S(t)$, $f(s, v) := -\beta sv + \gamma v$, which satisfies Assumption~\ref{as: equi}, and $M_1 = M_2 := 1$. Suppose
$S(t) = s(t) \to \bar{s}$ for all $(s(0), x(0)) \in \mathscr{I}$, then Theorem~\ref{thm: bound} says that
$\bar{s} \leq \frac{1}{\hat{G}(0)} = \frac{\gamma}{\beta} = \frac{1}{R_0}$,
which is aligned with known knowledge on the SIS model. In particular, when $R_0 > 1$, $\bar{s} \leq \frac{1}{R_0} = \frac{\gamma}{\beta} < 1$, and
hence $I(t) \to \bar{I} \geq 1 - \frac{\gamma}{\beta} > 0$. Thus, by Theorem~\ref{thm: bound}, we have $S(t) \to \frac{\gamma}{\beta}$, which
corresponds to the endemic state.
A further inspection reveals that as $S(t) = s(t) \to \bar{s}$, it holds that $\bar{s} = \frac{\gamma}{\beta}$ or $I(t) \to 0$. In the former case, the
LTI system $[G, \tilde{\Delta}]$ as described by \eqref{eq: CL_LTI} is given by
$\dot{I}(t) = \beta w(t)$,
which is an integrator and has a marginally stable mode at $0$. In the latter case,  the
LTI system $[G, \tilde{\Delta}]$ as described by \eqref{eq: CL_LTI} is given by
$\dot{I}(t) =(A+BKC)I(t)+\beta w(t)= (-\gamma+\beta)I(t)+\beta w(t)$,
whereby $\lambda (A+BKC)=-\gamma+\beta\le0$.
These are consistent with Theorem~\ref{thm: bound_iff}. \vspace{-3pt}

\subsection{SIR models} \vspace{-3pt}

The SIR model introduced in~\cite{KerMcK32} is given by   \vspace{-3pt}
\begin{align*}
  \dot{S}(t) &= -\beta S(t) I(t) \\
  \dot{I}(t) & = \beta S(t) I(t) - \gamma I(t) \\
  \dot{R}(t) & = \gamma I(t),
\end{align*}
where $S(t)$ denotes the proportion of the population that is susceptible at time $t$, $I(t)$ the proportion that is infected, $R(t)$ the proportion
that is removed or has recovered with immunity, $\beta$ the infection rate, $\gamma$ the recovery rate, and $S(0) + I(0) + R(0) = 1$. Define LTI
system $G$ as in \eqref{eq: LTI} with $x(t) := I(t)$, $A := -\gamma$, $B := 1$, $C := 1$, and nonlinear system $\Delta$ as in \eqref{eq:
  nonlinear} with $s(t) := S(t)$, $f(s, v) := -\beta sv$, which satisfies Assumption~\ref{as: equi0}, and $M_1 := 1$, $M_2 := \beta$. The set of initial
conditions of interest is $\mathscr{I} := \{(S, I) : S \geq 0, I > 0, S + I \leq 1\}$, which satisfies Assumption~\ref{as: initial_cond_strong}. \vspace{-3pt}

Observe that $S(t)$ is monotonically nonincreasing and bounded from below, so it converges as per the monotone convergence theorem \cite{rudin76}. Suppose
$S(t) \to \bar{s}>0$, whose continuity in $x(0)$ follows from Theorem~\ref{thm: s_value}, then Theorem~\ref{thm: bound_strict} states that
$\bar{s} < \frac{1}{M_2\hat{G}(0)} = \frac{\gamma}{\beta}=\frac{1}{R_0}$,
in which case $R_0 := M_2\hat{G}(0) = \frac{\beta}{\gamma}$. Moreover, Theorem~\ref{thm: s_value} may be applied to find $\bar{s}$ and
$\lim_{t \to \infty} R(t)$. This is consistent with the existing result~\cite[Theorem 2.1]{Het00}. It is noteworthy that the dynamics in $R$ are not
part of the feedback loop involving $S$ and $I$. Also observe that the LTI system $[G, \tilde{\Delta}]$ as described by \eqref{eq:
  autonomous_CL_LTI_strict} is given by
$\dot{I}(t) = (\beta \bar{s} - \gamma) I(t)$,
where $\beta \bar{s} - \gamma < 0$ is Hurwitz. This agrees with~Theorem~\ref{thm: bound_iff_strict}. \vspace{-3pt}

%%%%%%%%%%%%%%%%%%%%%%%%%%%%%%%%%%%%%%%%%%%%%%%%%%%%%%%
\begin{exmp}
Suppose $\beta=0.3$ and  $\gamma=0.2$, and let the initial conditions be chosen randomly in $\{(S,I,R) : S\ge 0, I>0, R\ge 0,
S+I=1\}$. It follows that $R_0=\frac{\beta}{\gamma}=1.5>1$.  The trajectories for $S(t)$ under  different  initial conditions are shown in
Figure \ref{fig:Ex_VB1}, which shows $\bar{s}<\frac{1}{R_0}$.

\begin{figure}
    \centering
    \includegraphics[scale=0.3]{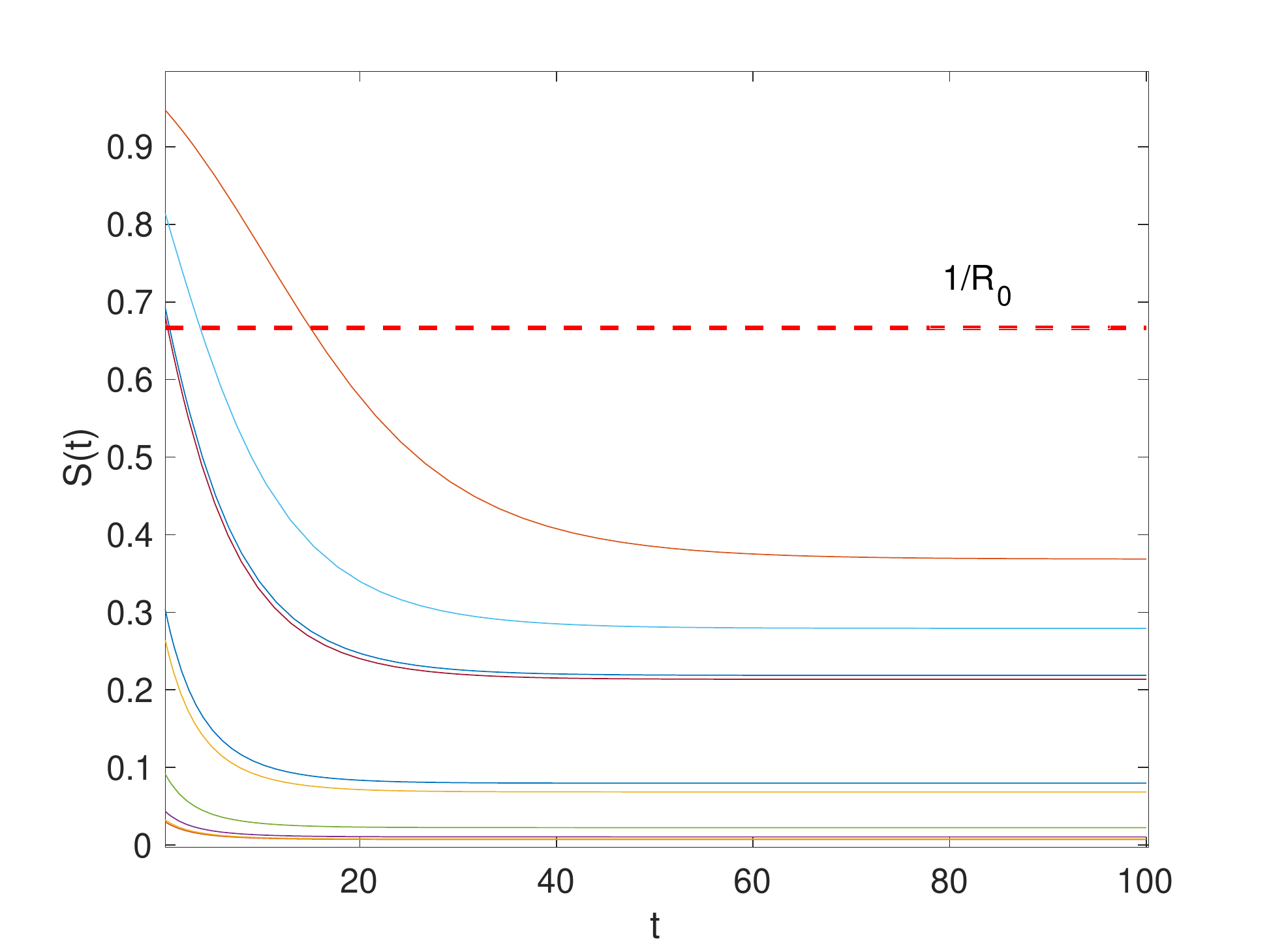}
    \centering \caption{Example of group SIR models: the trajectories of $S(t)$ under various initial conditions}
    \label{fig:Ex_VB1}
  \end{figure}

 \end{exmp} \vspace{-3pt}
%%%%%%%%%%%%%%%%%%%%%%%%%%%%%%%%%%%%%%%%%%%%%%%%%%%%%%%

The SIRS model with vital dynamics (balanced births and deaths)~\cite[Section 2.4]{Het00} is given by   \vspace{-3pt}
\begin{align} \label{eq: SIR_vital}
  \begin{split}
  \dot{S}(t) &= \mu -\mu S(t) -\beta S(t) I(t) + \delta R(t) \\
  \dot{I}(t) & = \beta S(t) I(t) - \gamma I(t) -\mu I(t) \\
  \dot{R}(t) & = \gamma I(t) - \mu R(t) - \delta R(t),
  \end{split}
\end{align}
where $\delta \geq 0$ is the rate at which immunity recedes following recovery and there is an inflow of newborns into the susceptible compartment at
rate $\mu > 0$ and deaths in all the compartments at rates $\mu S$, $\mu I$, and $\mu R$ respectively. Define LTI system $G$ as in \eqref{eq: LTI}
with $x(t) := [I(t), R(t)]$, $A := \STwoTwo{-(\gamma + \mu)}{0}{\gamma}{-(\delta+\mu)}$, $B := \STwoOne{\beta}{0}$, $C := I$, and nonlinear system
$\Delta$ as in \eqref{eq: nonlinear} with $s(t) := S(t)$, $f(s, v) := \mu - \mu s + \SOneTwo{-\beta s}{\delta} v$, which satisfies Assumption~\ref{as:
  equi}, and $M_1: = 1$, $M_2 := \SOneTwo{1}{0}$. The set of initial conditions of interest is
$\mathscr{I} := \{(S, I, R) : S \geq 0, I \geq 0, R \geq 0, S + I + R = 1\}$, which satisfies Assumption~\ref{as: initial_cond}. Global asymptotic stability
of the equilibrium of the SIR model with vital dynamics (where $\delta = 0$) has been shown in~\cite[Theorem 2.2]{Het00}
Suppose $S(t) \to \bar{s}$ for all $(s(0), x(0)) \in \mathscr{I}$ in \eqref{eq: SIR_vital}, Theorem~\ref{thm: bound} then
states that
$\bar{s} \leq \frac{1}{M_2\hat{G}(0)} = \frac{\gamma + \mu}{\beta}=\frac{1}{R_0}$,
where $R_0 := M_2\hat{G}(0) = \frac{\beta}{\gamma + \mu}$. In particular, when $R_0 > 1$, $\bar{s} < 1$. From \eqref{eq: SIR_vital}, this implies
that $I(t) \to \bar{I} > 0$. Thus, by Theorem~\ref{thm: bound},
$\bar{s} = \frac{\gamma + \mu}{\beta}$,
whereby $\bar{I} = \frac{(\delta+\mu)(\beta-\gamma-\mu)}{\beta(\delta+\gamma+\mu)}$, corresponding to the
endemic state. This is consistent with the existing result~\cite[Theorem 2.2]{Het00}, where $\delta$ is taken to be $0$. \vspace{-3pt}

%%%%%%%%%%%%%%%%%%%%%%%%%%%%%%%%%%%%%%%%%%%%%%%%%%%%%%%%%
\begin{exmp}
Suppose $\beta=0.3$, $\gamma=0.2$, $\delta=0.1$ and $\mu=0.001$, and let the initial conditions be chosen randomly in $\{(S,I,R) : S\ge 0, I>0, R\ge 0,
S+I=1\}$. It follows from the above results that $\bar{s}=\frac{\gamma+\mu}{\beta}=0.67$ and $\bar{I} =
\frac{(\delta+\mu)(\beta-\gamma-\mu)}{\beta(\delta+\gamma+\mu)}=0.11$.  The trajectories for $S(t)$ and $I(t)$ for different  initial conditions are shown in Figure \ref{fig:ExVB2}. It shows that $S(t)$ converges to the same value, $\bar{s}$, and $I(t)$ converges to $\bar{I}$ under various initial conditions. \vspace{-3pt}

\begin{figure}
    \centering
    \includegraphics[scale=0.2]{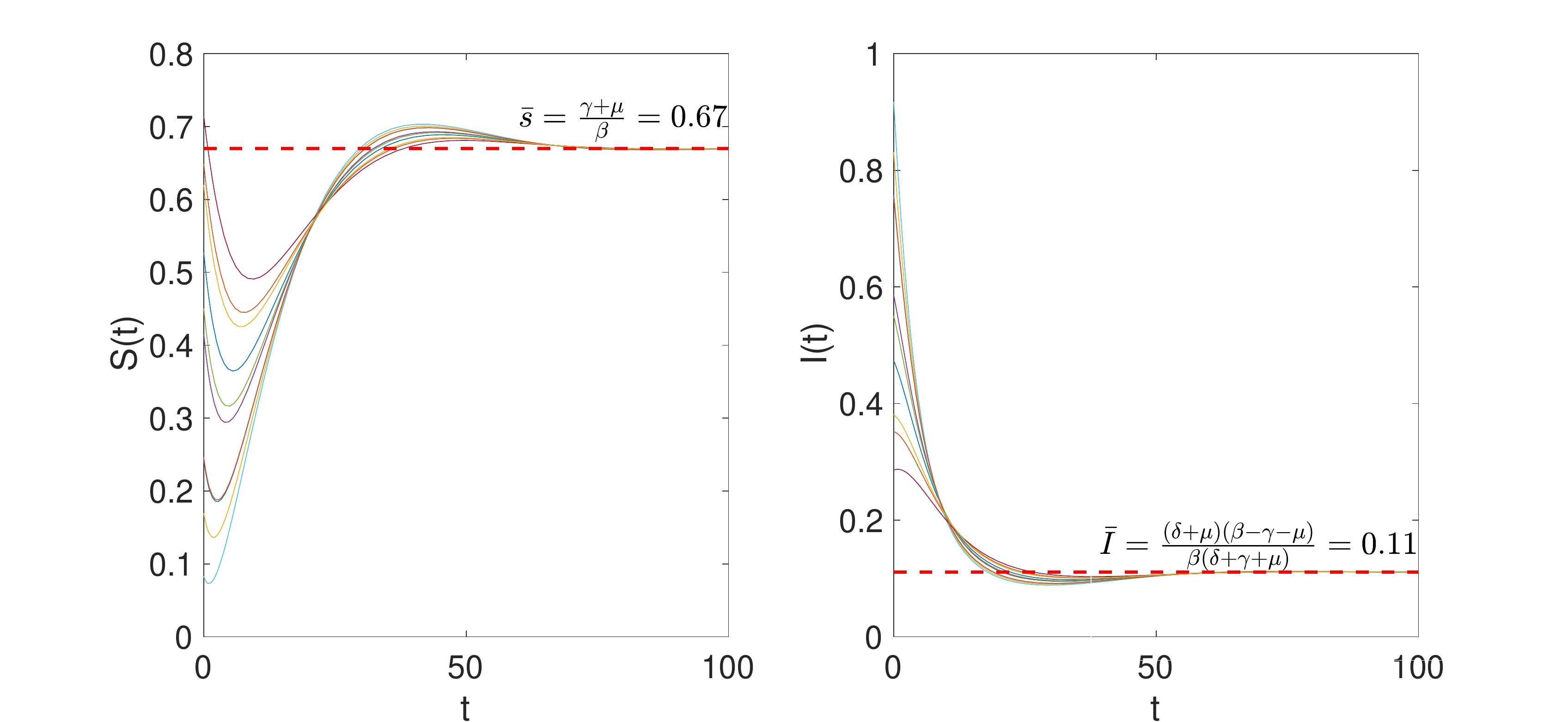}
    \centering \caption{Example of group SIR models with vital dynamics: the trajectories of $S(t)$ and $I(t)$ under various initial conditions}
    \label{fig:ExVB2}
  \end{figure}
  \end{exmp}
  %%%%%%%%%%%%%%%%%%%%%%%%%%%%%%%%%%%%%%%%%%%%%%%%%%%%%%%%%%5

\section{Networked epidemic models} \label{sec: networked_models} \vspace{-3pt}

Networked models capture the scenario where numerous groups or nodes are interconnected via a contact graph or interconnection network, defined by an
adjacency matrix $W = [w_{ij}]$. Each $w_{ij} \geq 0$ quantifies the strength of the connection from node $j$ to node $i$. Similarly to the group
epidemic models, each group/node in a network is made up of different compartments (susceptible, infected etc.) in a networked compartmental
model. It is worth noting that convergence analysis of networked epidemic models has not been completed in the literature to the authors' best
knowledge, except for the SIS model and those that are straightforwardly guaranteed by the monotone convergence theorem. \vspace{-3pt}

The \emph{basic reproduction number} (BRN) $R_0$ is a recurring threshold of interest in networked models, as is the case of group models. The
difference in this section from the last is that the BRN is given by the spectral radius of a nonnegative matrix here. \vspace{-3pt}

\subsection{SEIR models} \vspace{-3pt}

Suppose there are $N$ nodes. Let $s_i(t)$, $e_i(t)$, $p_i(t)$, and $r_i(t)$ denote the proportions of population that are susceptible, exposed,
infected, and removed, respectively, at time $t$ and at node $i \in \{1, \ldots, N\}$. The networked SEIR model~\cite[Section 3.3]{PBB20} is described
by   \vspace{-3pt}
\begin{align*}
  \dot{s}(t) &= -\diag(s(t)) [\diag(\beta_E) W e(t) +\diag(\beta_I) W p(t)] \\
  \dot{e}(t) & = \diag(s(t)) [\diag(\beta_E) W e(t) + \diag(\beta_I) W p(t)]  \\
  & \qquad\qquad\qquad\qquad\qquad\qquad\qquad\qquad - \diag(\sigma) e(t) \\
  \dot{p}(t) & = \diag(\sigma) e(t) - \diag(\gamma) p(t) \\
  \dot{r}(t) & = \diag(\gamma) p(t),
\end{align*}
where $\sigma$ denotes the transition rate from exposed to infected, $\beta_E$ and $\beta_I$ represent the transmission rates between susceptible and
exposed, and susceptible and infected, respectively. Define LTI system $G$ as in \eqref{eq: LTI} with $x(t) := [e(t)^T, p(t)^T]^T$,
$A := \STwoTwo{-\diag(\sigma)}{0}{\diag(\sigma)}{-\diag(\gamma)}$, $B := \STwoOne{I}{0}$, $C := I$, and nonlinear system $\Delta$ as in \eqref{eq:
  nonlinear} with $f(s, v) := -\diag(s)\SOneTwo{\diag(\beta_E)W}{\diag(\beta_I)W} v$, which satisfies Assumption~\ref{as: equi0}, and $M_1 = I$,
$M_2 = \SOneTwo{\diag(\beta_E)W}{\diag(\beta_I)W}$. The set of initial conditions of interest is
$\mathscr{I} := \{(s, e, p) : s \geq 0, e \geq 0, p \geq 0, s + p + e \leq 1_{n_s}, e + p > 0\}$, which satisfies Assumption~\ref{as:
  initial_cond_strong}.
Observe that each entry in $s(t)$ is monotonically nonincreasing and bounded from below, so by the monotone convergence theorem it converges. Suppose
$s(t) \to \bar{s}\gg 0$, whose continuity in $x(0)$ follows from Theorem~\ref{thm: s_value}, then Theorem~\ref{thm: bound_strict} says that there exists
$i$ such that   \vspace{-3pt}
\begin{align*}
 \bar{s}_i & < \frac{1}{\rho(M_2\hat{G}(0))} \\
           & = \frac{1}{\rho(\diag(\beta_E)W\diag(\sigma)^{-1} + \diag(\beta_I)W\diag(\gamma)^{-1})} \\
  & =: \frac{1}{R_0}.
\end{align*}
Theorem~\ref{thm: s_value} is applicable here for evaluating $\bar{s}$, $\lim_{t \to \infty} p(t)$ and $\lim_{t \to \infty} r(t)$. \vspace{-3pt}
%%%%%%%%%%%%%%%%%%%%%%%%%%%%%%%%%%%%%%%%%

\begin{exmp}\label{ex:SEIRnetwork}
  Consider a network consisting of 5 nodes depicted in Figure \ref{fig:networkexample}.
\begin{figure}
    \centering
    \includegraphics[scale=0.5]{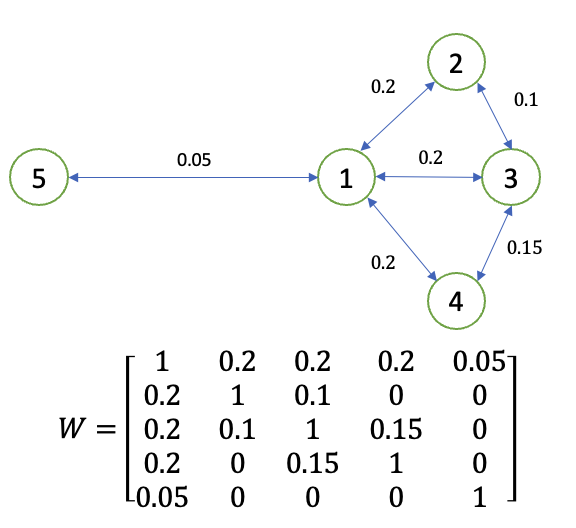}
    \centering \caption{ A contact graph and its adjacent matrix}
    \label{fig:networkexample}
\end{figure}
% Consider the same network in Figure \ref{fig:networkexample}
with $\beta_E=[0.2\;0.1\;0.1\;0.1\;0.05]^T$, $\beta_I=0.2\cdot[1\;1\;1\;1\;1]^T$,
$\sigma=0.1\cdot[1\;1\;1\;1\;1]^T$  and $\gamma=[0.3\;0.2\;0.1\;0.2\;0.1]^T$. 
It follows that $R_0=\rho(\diag(\beta_E)W\diag(\sigma)^{-1} + \diag(\beta_I)W\diag(\gamma)^{-1})=3.6987>1$. Suppose the epidemic is initiated by a
minority of the population in Node 5 being exposed to the disease, and the initial condition is given by $s(0)=[1\;1\;1\;1\;0.9999]^T, e(0)=[0\;0\;0\;0\;0.0001]^T, p(0)=r(0)=0$. The
trajectory of each component in $s(t)$ is shown in Figure \ref{fig:ExVIB1}, all of which
 converge to some value below $\frac{1}{R_0}$.  \vspace{-3pt}

\begin{figure}
    \centering
    \includegraphics[scale=0.25]{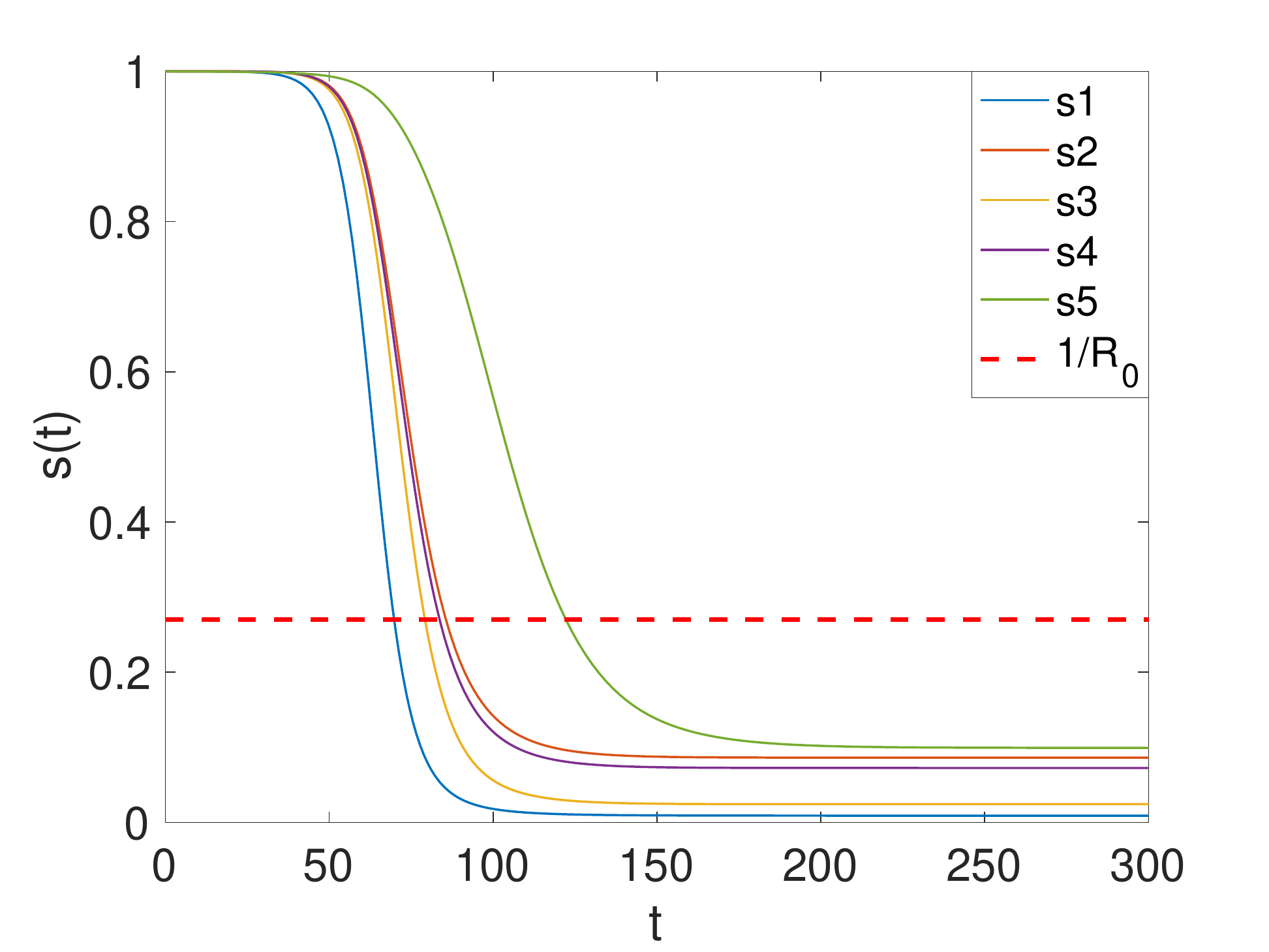}
    \centering \caption{Example of networked SEIR models: the trajectory of $s(t)$}
    \label{fig:ExVIB1}
  \end{figure}
\end{exmp}

%%%%%%%%%%%%%%%%%%%%%%%%%%%%%%%%%%%%%%%%
The networked SEIR model with vital dynamics is described by   \vspace{-3pt}
\begin{align*}
  \dot{s}(t) &= \mu - \diag(\mu)(s(t)) \\
  & \qquad\quad -\diag(s(t)) [\diag(\beta_E) W e(t) +\diag(\beta_I) W p(t)] \\
  \dot{e}(t) & = \diag(s(t)) [\diag(\beta_E) W e(t) + \diag(\beta_I) W p(t)]  \\
             & \qquad\qquad\qquad\qquad\qquad\quad - \diag(\sigma) e(t) - \diag(\mu)e(t)\\
  \dot{p}(t) & = \diag(\sigma) e(t) - \diag(\gamma) p(t) - \diag(\mu) p(t) \\
  \dot{r}(t) & = \diag(\gamma) p(t) - \diag(\mu) r(t).
\end{align*}
Define LTI system $G$ as in \eqref{eq: LTI} with $x(t) := [e(t)^T, p(t)^T]^T$,
$A := \STwoTwo{-\diag(\sigma + \mu)}{0}{\diag(\sigma)}{-\diag(\gamma + \mu)}$, $B := \STwoOne{I}{0}$, $C := I$, and nonlinear system $\Delta$ as in
\eqref{eq: nonlinear} with $f(s, v) := \mu - \diag(\mu)s -\diag(s)\SOneTwo{\diag(\beta_E)W}{\diag(\beta_I)W}v$, which satisfies Assumption~\ref{as:
  equi}, and $M_1 = I$, $M_2 = \SOneTwo{\diag(\beta_E)W}{\diag(\beta_I)W}$. Suppose $s(t) \to \bar{s}$ for all $(s(0), x(0)) \in \mathscr{I}$, then
Theorem~\ref{thm: bound} says that there exists $i$ such that
$\bar{s}_i \leq \frac{1}{\rho(M_2\hat{G}(0))}= \frac{1}{R_0}$,
where $R_0 := \rho(\diag(\beta_E)W\diag(\sigma + \mu)^{-1} + \diag(\beta_I)W\diag(\sigma)\diag(\sigma+\mu)^{-1}\diag(\gamma+\mu)^{-1})$. \vspace{-3pt}

%%%%%%%%%%%%%%%%%%%%%%%%%%%%%%%%%%%%%%%%
\begin{exmp}
Consider again Example \ref{ex:SEIRnetwork} but with vital dynamics and let $\mu=0.001\cdot[1\;1\;1\;1\;1]^T$. 
It follows that $R_0=\rho(\diag(\beta_E)W\diag(\sigma + \mu)^{-1}+
\diag(\beta_I)W\diag(\sigma)\diag(\sigma+\mu)^{-1}\diag(\gamma+\mu)^{-1})=3.6482>1$. It can be verified by simulation that all initial conditions in
$\mathscr{I}$ will lead to the same $\bar{s}$.  Let the initial condition be given also by $s(0)=[1\;1\;1\;1\;0.9999]^T, e(0)=[0\;0\;0\;0\;0.0001]^T, p(0)=r(0)=0$. The
trajectory of each component in $s(t)$ is shown in Figure \ref{fig:ExVIB2}. In particular, observe that $\bar{s}_1<\frac{1}{R_0}$.  \vspace{-3pt}

\begin{figure}
    \centering
    \includegraphics[scale=0.25]{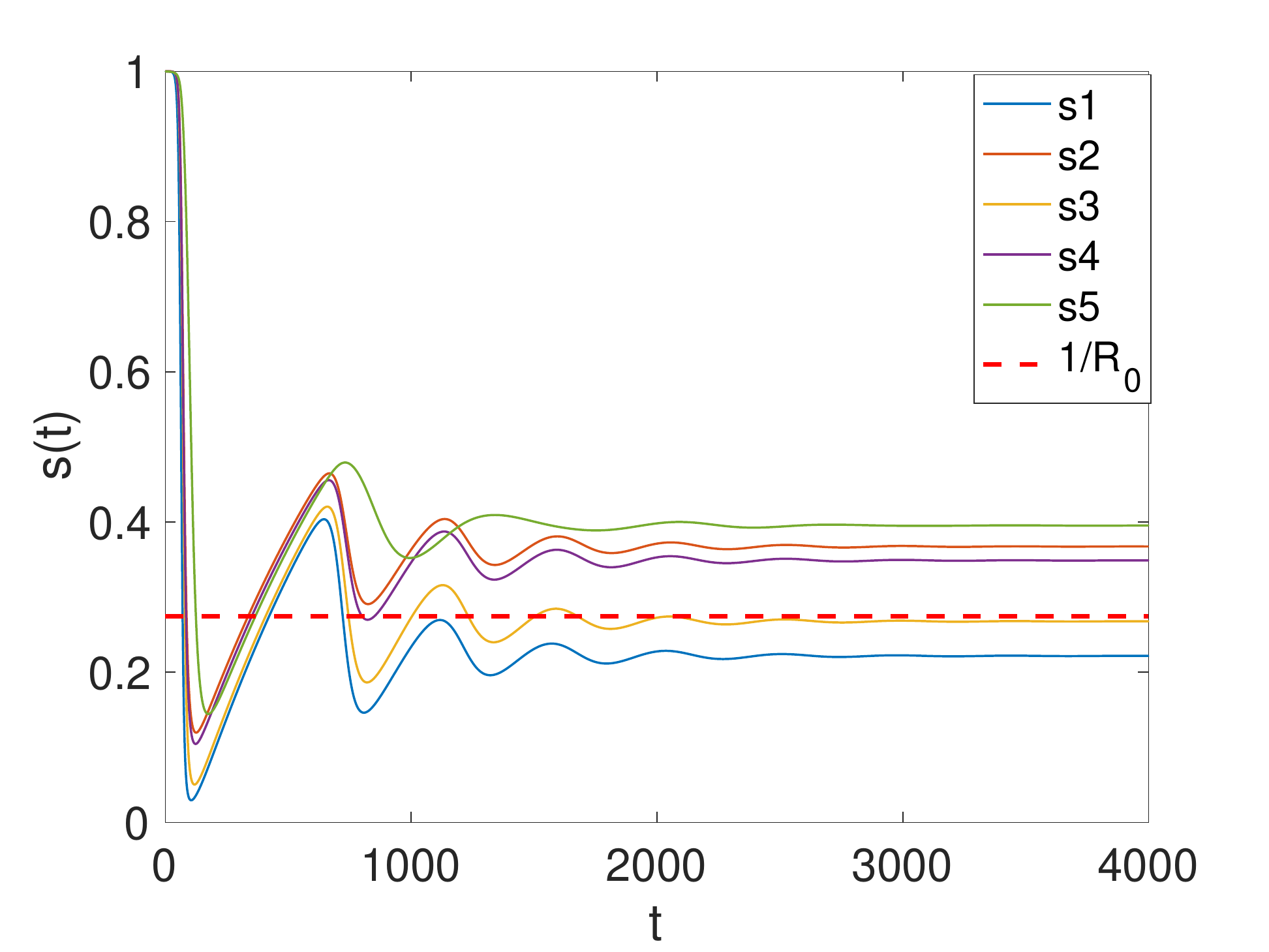}
    \centering \caption{Example of networked SEIR models with vital dynamics: the trajectory of $s(t)$}
    \label{fig:ExVIB2}
  \end{figure}
\end{exmp}

%%%%%%%%%%%%%%%%%%%%%%%%%%%%%%%%%%%%%%%%

% Note that Theorem~\ref{thm: bound} is not suited to studying networked SEIRS models for the same reason detailed in the subsection on networked SIR models.

\subsection{SAIR models} \vspace{-3pt}

The networked SAIR model~\cite[Section 3.3]{PBB20} is given by   \vspace{-3pt}
\begin{align*}
  \dot{s}(t) &= -\diag(\beta_A)\diag(s(t)) W a(t) \\
  & \qquad\qquad\qquad\qquad - \diag(\beta_I) \diag(s(t)) W p(t) \\
  \dot{a}(t) & = \diag(\beta_A) Q \diag(s(t)) W a(t) \\
  & \qquad + \diag(\beta_I) Q \diag(s(t)) W p(t) - \diag(\sigma + \kappa) a(t) \\
  \dot{p}(t) & = \diag(\beta_A) (I - Q)\diag( s(t)) W a(t) \\
  & \qquad\qquad\qquad + \diag(\beta_I) (I - Q) \diag(s(t)) W p(t) \\
  & \qquad \qquad \qquad\qquad\quad  + \diag(\sigma) a(t) - \diag(\gamma + \nu) p(t) \\
  \dot{r}(t) & = \diag(\kappa) a(t) + \diag(\gamma) p(t) \\
  \dot{d}(t) & = \diag(\nu) p(t),
\end{align*}
where $a(t)$ represents the proportion of the population at time $t$ that has caught the disease but is asymptomatic, $d(t)$ the proportion that
is deceased, $\beta_A$ and $\beta_I$ the infection rates between susceptible and asymptomatic-infected, and susceptible and infected-symptomatic
individuals respectively, and $\sigma$ the progression rate from asymptomatic $a$ to symptomatic infected $p$, $\kappa$ and $\gamma$ the recovery
rates for $a$ and $p$, respectively, $\nu$ the progression rate from infected $p$ to deceased $d$, and $q$ and $1-q$ the probabilities or proportions
of susceptible individuals transitioning from $s$ to $a$ and $p$ respectively, and $Q = \diag(q)$. \vspace{-3pt}

Define LTI system $G$ as in \eqref{eq: LTI} with $x(t) := [a(t)^T, p(t)^T]^T$,
$A := \STwoTwo{-\diag(\sigma + \kappa)}{0}{\diag(\sigma)}{-\diag(\gamma + \nu)}$, $B := \STwoOne{Q}{I-Q}$, $C := I$, and nonlinear system $\Delta$
as in \eqref{eq: nonlinear} with $f(s, v) := -\diag(s)\SOneTwo{\diag(\beta_A)W}{\diag(\beta_I)W} v$, which satisfies Assumption~\ref{as: equi0}, and
$M_1 = I$, $M_2 := \SOneTwo{\diag(\beta_A)W}{\diag(\beta_I)W}$. The set of initial conditions of interest is
$\mathscr{I} := \{(s, a, p) : s \geq 0, a \geq 0, p \geq 0, s + a + p \leq 1_{n_s}, a + p > 0\}$, which satisfies Assumption~\ref{as: initial_cond_strong}.
Observe that every entry in $s(t)$ is monotonically nonincreasing and bounded from below, whereby it converges by the monotone convergence theorem. Suppose
$s(t) \to \bar{s}\gg 0$, whose continuity in $x(0)$ follows from Theorem~\ref{thm: s_value}, Theorem~\ref{thm: bound_strict} then states that there exists $i$ such that
$\bar{s}_i < \frac{1}{\rho(M_2\hat{G}(0))} = \frac{1}{R_0}$,
in which   \vspace{-3pt}
\begin{align*}
  R_0 &: = \rho(\diag(\beta_A)W \diag(\kappa + \sigma)^{-1}Q \\
      & \quad + \diag(\beta_I)W \diag(\sigma)\diag(\gamma + \nu)^{-1}\diag(\kappa + \sigma)^{-1}Q \\
  & \quad + \diag(\beta_I)W\diag(\gamma + \nu)^{-1}(I - Q)).
\end{align*}
Note that Theorem~\ref{thm: s_value} is applicable here for computing $\bar{s}$, $\lim_{t \to \infty} r(t)$, and $\lim_{t \to \infty} d(t)$.  \vspace{-3pt}

%%%%%%%%%%%%%%%%%%%%%%%%%%%%%%%%%%%%%% 

\begin{exmp}
Consider the same network in Figure \ref{fig:networkexample} with $\beta_A=[0.2\;0.1\;0.1\;0.1\;0.05]^T$, $\beta_I=0.1\cdot[1\;1\;1\;1\;1]^T$, $q=0.1\cdot[1\;1\;1\;1\;1]^T$,
$\sigma=0.3\cdot[1\;1\;1\;1\;1]^T$, $\gamma=\kappa=[0.2\;0.1\;0.1\;0.1\;0.05]^T$ and $\nu=0.001\cdot[1\;1\;1\;1\;1]^T$. 
It follows from the preceding result that $R_0=1.9488>1$. Suppose the epidemic is initiated by a
minority of the population in Node 5 being asymptomatic-infected, and the initial condition is given by $s(0)=[1\;1\;1\;1\;0.99]^T,
a(0)=[0\;0\;0\;0\;0.01]^T, p(0)=r(0)=d(0)=0$. As shown in Figure \ref{fig:ExVIC}, $\bar{s}_5$
is less than $\frac{1}{R_0}$.    \vspace{-3pt}

\begin{figure}
    \centering
    \includegraphics[scale=0.2]{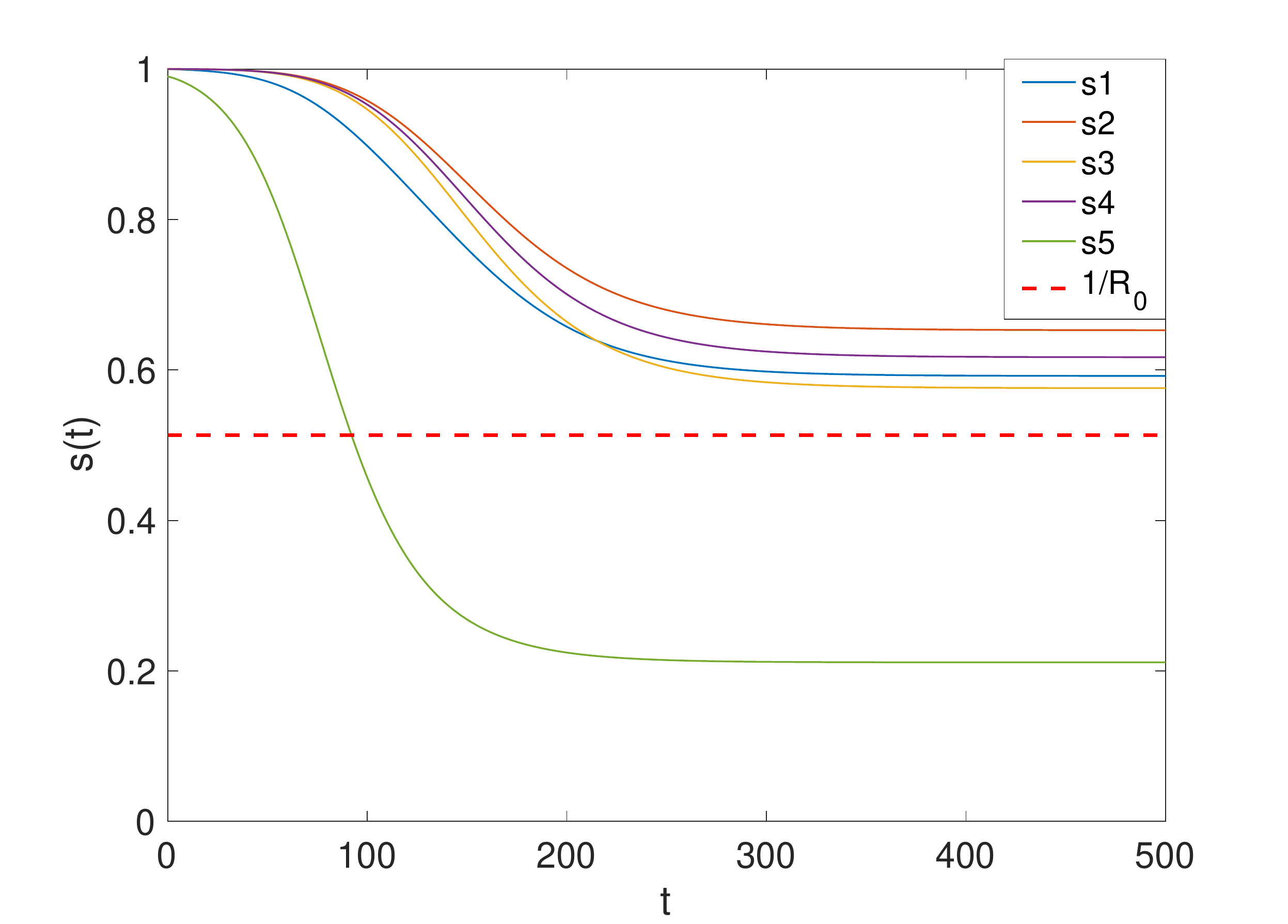}
    \centering \caption{Example of networked SAIR models: the trajectory of $s(t)$}
    \label{fig:ExVIC}
  \end{figure}
\end{exmp}
%%%%%%%%%%%%%%%%%%%%%%%%%%%%%%%%%%%%%

\section{Conclusion} \label{sec: con}   \vspace{-3pt}

We developed two positive feedback system frameworks for the steady-state analysis of epidemic models and showed that the reciprocal of the basic
reproduction number quantifies the level of penetration into at least one subgroup in a networked epidemic model. Two significantly different
scenarios involving the existence and nonexistence of the endemic state were considered, and they were shown to correspond to distinct dynamics in the
positive feedback system. In the case where there is no endemic state, formulae for computing the convergence limits in the epidemic models were also
provided. Various illustrative examples on different compartmental epidemic models were studied and simulated to validate our results. \vspace{-3pt}

Interesting future research directions include investigating with similar approaches the discrete-time models~\cite{PLBKB19}, the control aspects in
epidemic models~\cite{NPP16, RamMar17, YLAC21}, and the competitive propagation of more than one virus~\cite{LPNTBB19}. Furthermore, one may
investigate the effects of the changes in the networks and infection/recovery rates on the BRNs. These changes may arise from public health policies
on quarantine, isolation, social distancing, mask mandates, and/or vaccinations. \vspace{-3pt}

\bibliographystyle{plain} 
\bibliography{refs}

% \begin{IEEEbiography} {Sei Zhen Khong} received the Bachelor of Electrical Engineering degree (with first class honours) and the Ph.D. degree from
%   University of Melbourne, Australia, in 2008 and 2012, respectively. He has held research positions at the Department of Electrical and Electronic
%   Engineering, University of Melbourne, Australia, Department of Automatic Control, Lund University, Sweden, Institute for Mathematics and its
%   Applications, University of Minnesota Twin Cities, USA, and Department of Electrical and Electronic Engineering, University of Hong Kong, China. His
%   research interests include network control, robust control, systems theory, and extremum seeking.
% \end{IEEEbiography}

% \begin{IEEEbiography}
%   {Lanlan Su} is a Lecturer at the Department of Automatic Control and Systems Engineering, University of Sheffield, UK. She received the B.E. degree
%   in Electrical Engineering from Zhejiang University, China, in 2014, and the Ph.D. degree in Control Engineering from The University of Hong Kong, in
%   2018. From 2019 to 2022, she was a Lecturer at the School of Engineering, University of Leicester, UK. Before that, she was a postdoctoral research
%   associate at University of Notre Dame, USA. She is an awardee of the Hong Kong Ph.D. Fellowship Scheme established by the Research Grants Council of
%   Hong Kong. Dr. Su’s research interests include robust control, networked control systems and distributed systems.
% \end{IEEEbiography}
\end{document}